\documentclass[final,3p,times]{elsarticle}%
\usepackage{amsfonts}
\usepackage{amsmath}
\usepackage{amssymb}
\usepackage[english]{babel}
\usepackage{graphicx}
\usepackage{subfig}
\usepackage{placeins}
\usepackage{booktabs}
\usepackage{dcolumn}
\usepackage{array}
\usepackage{longtable}%
\providecommand{\U}[1]{\protect\rule{.1in}{.1in}}
\makeatletter
\def\ps@pprintTitle{ \let\@oddhead\@empty
 \let\@evenhead\@empty
 \def\@oddfoot{} \let\@evenfoot\@oddfoot}
\makeatother

\begin{document}

\title{Quantum cosmology in an anisotropic n-dimensional universe}

\author{F. A. P. Alves-J\'{u}nior} 
\author{M. L. Pucheu}\ead{mlaurapucheu@fisica.ufpb.br}\address{Departamento de F\'{i}sica, Universidade Federal da Para\'{i}ba, Caixa Postal 5008, 58051-900, Jo\~ao Pessoa-PB, Brazil.}
\author{A. B. Barreto}\ead{adrianobraga@fisica.ufpb.br} \address{Instituto Federal de Educação, Ciência e Tecnologia do Rio Grande do Sul, 95043-700, Caxias do Sul-RS, Brazil} \address{Departamento de F\'{i}sica, Universidade Federal da Para\'{i}ba, Caixa Postal 5008, 58051-900, Jo\~ao Pessoa-PB, Brazil.}
\author{C. Romero}\ead{cromero@fisica.ufpb.br} \address{Departamento de F\'{i}sica, Universidade Federal da Para\'{i}ba, Caixa Postal 5008, 58051-900, Jo\~ao Pessoa-PB, Brazil.}

\begin{abstract}
We investigate quantum cosmological models in an n-dimensional anisotropic
universe in the presence of a massless scalar field. Our basic inspiration
comes from Chodos and Detweiler's classical model which predicts an
interesting behaviour of the extra dimension, shrinking down as time goes by.
We work in the framework of a \ recent geometrical scalar-tensor theory of
gravity. Classically, we obtain two distinct type of solutions. One of them
has an initial singularity while the other represents a static universe
considered as a whole. By using the canonical approach to quantum cosmology,
we investigate how quantum effects could have had an influence in the past
history of these universes.
\end{abstract}

\begin{keyword}
Extra dimensions \sep Weyl geometry \sep quantum cosmology
\end{keyword}

\maketitle

\section{Introduction}

In the last decades a great deal of work has gone into scalar-tensor theories
of gravity, particularly in the context of inflationary models and also in
attempts to explain the observed acceleration of the universe. In all these,
the scalar field plays an essential role, although its nature and origin as
yet remains unclear. However, in a recently proposed scalar-tensor theory, the
nature of the scalar field is attributed to the space-time geometry
\cite{Almeida}. In this picture, physical and geometrical objects are, by
construction, invariant under a new group of symmetry, namely, the group of
Weyl transformations, and this leads to a natural mapping between the action
of a scalar-tensor theory with a non-minimally coupled scalar field in a
non-Riemannian space-time and the action of general relativity with a massless
scalar field coupled to gravity through a dimensionless parameter. Recent
applications of this new theoretical proposal to cosmology include scenarios
displaying unusual geometrical space-time behaviour \cite{Pucheu}.

Among other alternative approaches to gravity theories, in which the scalar
field emerges, we would like to call attention for the modern $n$-dimensional
models of the universe. These have been developed in many different contexts,
starting from the seminal Kaluza-Klein ideas to string cosmology
\cite{string}. Even in a purely classical general relativistic framework a
particular appealing cosmological model worth of mentioning is the one
obtained in general relativity by Chodos and Detweiler, who put forward the
idea that the present stage of the Universe evolved from a five-dimensional
scenario in which the extra dimension becomes unobservably small due to a kind
of dynamical contraction \cite{ChodosPhysRevD21}. Following the same
direction, \ other higher-dimensional general anisotropic models have been
considered also in scalar-tensor theories of gravity \cite{SThigherdim}.

The introduction of scalar fields and higher dimensions are also motivated by
the attempt to answer many open questions in classical cosmology, particularly
those related to the early phases of the universe. One possibility of
examining these questions in a deeper way is to go beyond the classical level
and look for a new picture in which quantum effects are taken into account. An
important contribution to this line of research has been provided by the
quantum cosmology program \cite{QCapproach}. It should be said, however, that
there are currently many technical and conceptual difficulties with this
approach. For instance, a well known problem in quantum cosmology is the
definition of time, a problem often referred to as \textit{the problem of
time} \cite{problemoftime}. Indeed, it turns out that quantum cosmology does
not specify in a unique way a parameter that plays the role of time. In the
general relativistic context, there have been several attempts to overcome
this difficulty. A well known way of tackling the problem is by introducing
matter content into the model, the latter usually being represented by a
scalar field associated to a fluid with a barotropic equation of state
\cite{Barotropic}. Another interesting attempt to find a possible solution to
the problem of time in the framework of Brans-Dicke theory was given recently
\cite{FarajollahiIntJTheorPhys49}, in which there is no need to add matter in
the form of a scalar field as the gravitational theory itself provides such a
field \footnote{The quantization of Brans-Dicke theory of gravity has also
been considered in a standard way using the Schutz's formalism
\cite{BDquantisation}}. By choosing suitable canonical transformations, the
Brans-Dicke scalar field may be identified with time in the sense of the usual
Schr\"{o}dinger picture. By the same token, in the quantization of a
geometrical scalar-tensor theory we can naturally relate the intrinsic scalar
field to a parameter that measures the evolution of the system at the quantum level.

The goal of the present work is to analyse quantum cosmological scenarios
predicted by the geometrical scalar-tensor theory in an anisotropic
n-dimensional space-time. In the context of general relativity, a similar
problem was recently considered by P. Letelier \cite{LetelierPhysRevD82}. The
paper is organized as follows. We begin, in Section 2, with a brief review of
the basic tenets of the geometrical scalar-tensor gravitational theory. In
Section \ref{SecClassModel}, we present the classical solutions of the
n-dimensional model in the light of the Lagrangian formalism. We then proceed
to perform the Hamiltonian formalism and propose some canonical
transformations to decouple the canonical variables. We check that the
solutions obtained from the Lagrangian formalism are also solutions of the
Hamiltonian equations. Next, in Section \ref{ModelQuant}, we carry out the
canonical quantization of the model. By assuming that the classical geometry
has a flat spatial section we obtain the wave function of the universe and
calculate the expectation values according to the many-worlds interpretation. Finally, in
Section \ref{FinalRemarks}, we discuss our results.

\section{The geometrical gravitational theory}

\label{intro}

Let us begin by considering the gravitational sector of the non-minimally
coupled scalar-tensor action
\begin{equation}
\mathcal{S}=\int d^{n}x\sqrt{|g|}\left[  e^{-\phi}\left(  R+\omega g^{\mu\nu
}\phi_{,\mu}\phi_{,\nu}\right)  -V(\phi)\right]  ,
\label{lagrangian weyl frame}%
\end{equation}
defined on a $n$-dimensional space-time \footnote{This action can be regarded
as the n-dimensional generalization of the Jordan-Brans-Dicke action
\cite{Brans1961}.}, with $R$ denoting the $n$-dimensional curvature scalar
\footnote{We shall adopt the following definition of the curvature tensor:
$R_{\,\beta\mu\nu}^{\alpha}=\Gamma_{\beta\mu,\nu}^{\alpha}-\Gamma_{\beta
\nu,\mu}^{\alpha}+\Gamma_{\beta\mu}^{\sigma}\Gamma_{\sigma\nu}^{\alpha}%
-\Gamma_{\beta\nu}^{\sigma}\Gamma_{\sigma\mu}^{\alpha}$. The Ricci tensor is
defined as $R_{\mu\nu}=R_{\mu\alpha\nu}^{\alpha}$.}, $g$ the determinant of
the metric tensor $g_{\mu\nu}$, and $\omega$ being a dimensionless parameter.
As in the four-dimensional case, the field equations for $g_{\mu\nu}$ and
$\phi$, together with the non-metricity condition that characterizes a Weyl
integrable space-time (WIST), are easily obtained by applying the Palatini's
variational method to the above action (See ref. \cite{Almeida}). Thus, the
variation of (\ref{lagrangian weyl frame}) with respect to the affine
connection leads to
\begin{equation}
\nabla_{\alpha}g^{\mu\nu}=-\frac{2}{n-2}\phi_{,\alpha}\;g^{\mu\nu},
\label{nonmetricity cond}%
\end{equation}
where $\phi_{,\alpha}=\partial_{\alpha}\phi$. This is precisely the
non-metricity condition mentioned above, and that, in a certain sense, leads,
from first principles, to the determination of the space-time geometry
\cite{Weylgeometry} . From the above $\psi=\frac{2}{n-2}\phi$ plays the role
of the n-dimensional Weyl scalar field \footnote{Let us recall that
Eq.(\ref{nonmetricity cond}) gives an expression for the Weylian affine
connection in terms of the two fundamental geometrical elements of the
manifold, namely, the metric tensor and the scalar field. This may be written
as $\Gamma_{\mu\nu}^{\alpha}=\left\{  _{\mu\nu}^{\alpha}\right\}  -\frac{1}%
{2}g^{\alpha\beta}\left(  g_{\mu\beta}\psi_{,\nu}+g_{\beta\nu}\psi_{,\mu
}-g_{\mu\nu}\psi_{,\beta}\right)  $, with $\left\{  _{\mu\nu}^{\alpha
}\right\}  $ denoting the Christoffel symbols and $\psi$, the geometric scalar
field.}. In the terminology of the geometrical scalar-tensor theory, a Weyl
frame is the set $(M,g,\psi)$ characterized by the metric tensor $g$ and the
scalar field $\psi$ defined on the manifold $M$. An important property of the
Weyl geometry is that the non-metricity condition $\nabla_{\alpha}g^{\mu\nu
}=-\psi_{,\alpha}g^{\mu\nu}$ is invariant under the set of transformations
\begin{align}
\bar{g}_{\mu\nu}  &  =e^{f}g_{\mu\nu},\label{weyl transformations}\\
\bar{\psi}  &  =\psi+f.\nonumber
\end{align}
That is, in the new frame $(M,\bar{g},\bar{\psi})$ we have $\nabla_{\alpha
}\overline{g}^{\mu\nu}=-\bar{\psi}_{,\alpha}\overline{g}^{\mu\nu}$. Clearly,
these transformations preserve the geodesic curves, since the affine
connection is kept invariant. Because $\bar{g}_{\mu\nu}$ and $g_{\mu\nu}$ are
related by a conformal transformation the causal structure these metric define
on the manifold $M$ does not change when we go from one Weyl frame to another.
By setting $f=-\frac{2}{n-2}\phi=-\psi$ in (\ref{weyl transformations}) we
have $\bar{\psi}=0$. Because we recover the Riemannian compatibility condition
between the metric and affine connection, this frame is usually called the
\textit{Riemann frame}, and is denoted as the set $(M,\bar{g},0)$.

It is not difficult to verify that in the Riemann frame $(M,\bar{g},0)$ the
action (\ref{lagrangian weyl frame}) becomes
\begin{equation}
\mathcal{\bar{S}}=\int d^{n}x\sqrt{|\bar{g}|}\left[  \bar{R}+\omega\bar
{g}^{\mu\nu}\phi_{,\mu}\phi_{,\nu}-e^{\frac{n}{n-2}\phi}V(\phi)\right]  ,
\label{labrangian riemann frame}%
\end{equation}
which, for $\omega=\frac{1}{2}$, is formally identical to the $n$-dimensional
Hilbert-Einstein action of a scalar field minimally coupled with gravity with
a potential $U(\phi)$ given by $U(\phi)=e^{\frac{n}{n-2}\phi}V(\phi)$. In
fact, the analogy between the two configurations is even more apparent if we
recall that in the Riemann frame, particles and light rays will follow
Riemannian metric and affine geodesics, respectively. In the next section, we
shall investigate the cosmological scenarios that are generated by the action
(\ref{labrangian riemann frame}), when we take $V(\phi)=0$.

\section{The Classical Cosmological Model}

\label{SecClassModel}

\subsection{The Lagrangian formalism}

\label{subsec2.1}

We shall now consider the n-dimensional, $n>4$, anisotropic cosmological model
whose geometry is described by the following line element
\begin{equation}
ds^{2}=N(t)^{2}dt^{2}-a(t)^{2}\left(  dx^{2}+dy^{2}+dz^{2}\right)
-b(t)^{2}\sum_{i=1}^{n-4}{dl_{i}}^{2}, \label{ds-model}%
\end{equation}
with $N(t)$ denoting the lapse function, $a(t)$ being the scale factor
associated with the usual three spatial dimensions, and $b(t)$ representing
the scale factor of the $(n-4)$-dimensions, the latter being assumed to be compact.

\begin{figure}[tbh]
\centering
\includegraphics[scale=0.7]{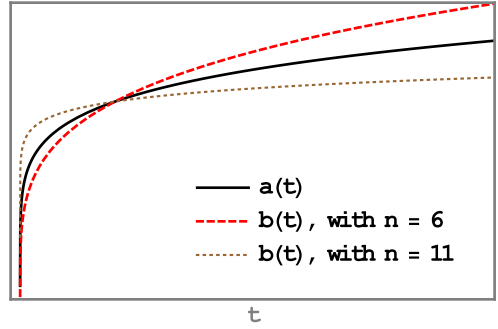} \caption{Scale factors in
(\ref{class sol 1})}%
\label{PlotEq9}%
\end{figure}

\begin{figure}[tbh]
\centering
\subfloat[]{
\includegraphics[scale=0.7]{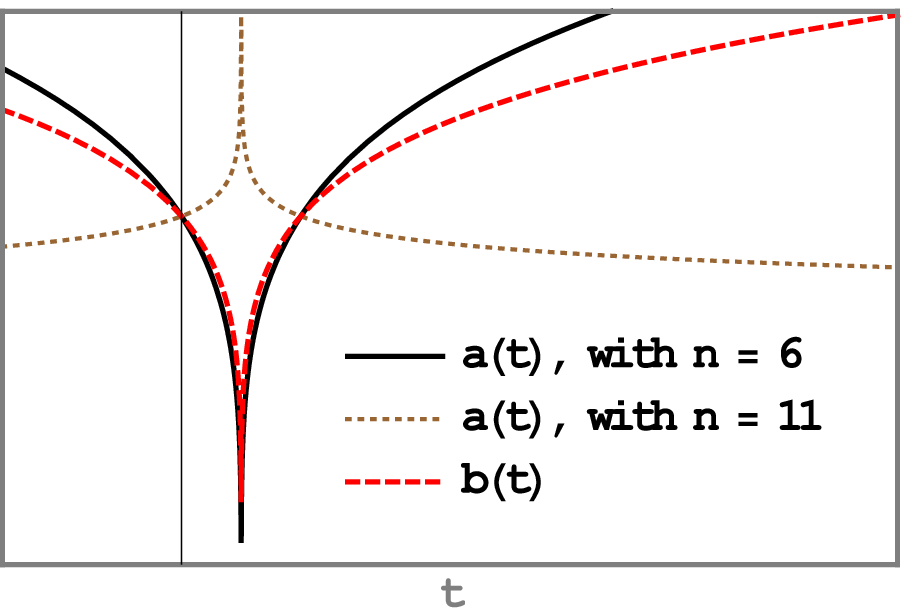}} \quad\subfloat[]{
\includegraphics[scale=0.7]{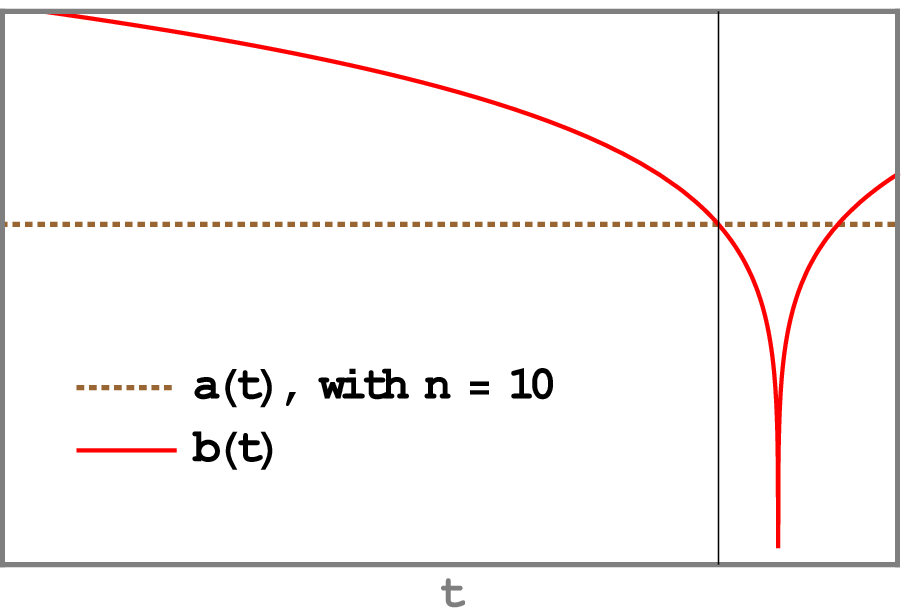}}\caption{Scale factors in
(\ref{class sol 2})}%
\label{PlotEq10}%
\end{figure}

The reduced action corresponding to (\ref{labrangian riemann frame}) written
in terms of the geometry given by the line element (\ref{ds-model}) takes the
following form:
\begin{equation}
\mathcal{S}_{\text{red}}=V_{o}\int dt\left[  -\frac{6}{N}\dot{a}^{2}%
ab^{n-4}-\frac{6(n-4)}{N}\dot{b}\dot{a}a^{2}b^{n-5}-\frac{(n-4)(n-5)}{N}%
\dot{b}^{2}a^{3}b^{n-6}+\frac{\omega}{N}a^{3}b^{n-4}\dot{\phi}^{2}\right]  ,
\label{red-action}%
\end{equation}
where the over dot denotes differentiation with respect to the time coordinate
$t$, while $V_{o}$ stands for the integration on the $(n-1)$-dimensional space
defined by the compact extra dimensions \footnote{In the derivation of reduced
action we have dropped surface terms, which do not contribute to the field
equations.}. From (\ref{red-action}) we write the Lagrangian of the model as
\begin{equation}
L\equiv-\frac{6}{N}\dot{a}^{2}ab^{n-4}-\frac{6(n-4)}{N}\dot{b}\dot{a}%
a^{2}b^{n-5}-\frac{(n-4)(n-5)}{N}\dot{b}^{2}a^{3}b^{n-6}+\frac{\omega}{N}%
a^{3}b^{n-4}\dot{\phi}^{2}. \label{Lagrang}%
\end{equation}
Now, if we set $N(t)\equiv1$, the field equations, obtained from the
Euler-Lagrange equations, are
\begin{align}
3H_{a}^{2}+3(n-4)H_{a}H_{b}+\frac{(n-4)(n-5)}{2}H_{b}^{2}  &  =\frac{\omega
}{2}\dot{\phi}^{2},\nonumber\label{eq of motion}\\
2\dot{H}_{a}+3H_{a}^{2}+(n-4)\dot{H}_{b}+2(n-4)H_{a}H_{b}+\frac{(n-3)(n-4)}%
{2}H_{b}^{2}  &  =-\frac{\omega}{2}\dot{\phi}^{2},\\
3\dot{H}_{a}+6H_{a}^{2}+(n-5)\dot{H}_{b}+3(n-5)H_{a}H_{b}+\frac{(n-4)(n-5)}%
{2}H_{b}^{2}  &  =-\frac{\omega}{2}\dot{\phi}^{2},\nonumber\\
3H_{a}\dot{\phi}+(n-4)H_{b}\dot{\phi}+\ddot{\phi}  &  =0,\nonumber
\end{align}
where we are defining $H_{a}=\frac{\dot{a}}{a}$ and $H_{b}= \frac{\dot{b}}{b}%
$. A solution of the equations of motion above is given by the following set
\begin{align}
\label{class sol 1}a(t)=a_{0} |C(t-t_{0}) -1|^{\frac{1}{6}}, \qquad b(t)=b_{0}
|C( t-t_{0})-1 |^{\frac{1}{2(n-4)}}, \qquad\phi(t)=\phi_{0} \pm\sqrt{\frac
{1}{\omega}\left( \frac{2}{3}+\frac{n-5}{4(n-4)}\right) }\; \ln|C( t-t_{0})-1
|,
\end{align}
with $a_{0}$, $b_{0}$, $\phi_{0}$ and $C$ denoting integration constants.
It is not difficult to verify that the solutions (\ref{class sol 1}) represent
a universe in which both the usual three dimensions and the extra $n-4$
dimensions expands as time passes, with a space-time singularity at
$t=t_{0}+1/C$ (See, Fig. 1 below) \ref{PlotEq9}. In this solution, since the
scalar field is a real function of time, it is required that $\omega>0$.
\FloatBarrier\FloatBarrier
On the other hand, a set of distinct solutions is given by
\begin{align}
\label{class sol 2}a(t)=a_{0}|C(t-t_{0})-1|^{\frac{(10-n)}{18}}, \qquad
b(t)=b_{0}|C(t-t_{0}) -1|^{\frac{1}{6}}, \qquad\phi(t)=\phi_{0} \pm\frac{1}%
{6}\sqrt{\frac{-n^{2}+17n+20}{3 \omega}}\;\ln|C(t-t_{0}) -1|.
\end{align}
As in the previous solution (\ref{class sol 1}), the behaviour of the scale
factor of the extra dimensions (\ref{class sol 2}) leads to a singularity as
$t\rightarrow t_{0}+1/C$. There are, however, some differences in this case.
While the extra dimensions always expand, the behaviour of the $3$-dimensional
spatial dimensions depends on the dimensionality of the model: If $n<10$, then
they start from a singularity at $t=t_{0}$ and expand forever; if $n>10$, they
undergo indefinitely a contraction phase; and if $n=10$, they remain constant
as $a(t)=a_{0}=const$. In Figure \ref{PlotEq10}, we show the behaviour of both
scale factors, $a(t)$ and $b(t),$ for $n=6$ and $n=11$. Let us also note that
$\omega$ must be positive for $n<19,$ and negative for $n\geq19$.

\FloatBarrier
\FloatBarrier

Here, it is interesting to note that according \ to (\ref{class sol 2}), a
curious scenario arises when $n=10$. In that case, the $3$-dimensional scale
factor $a$ is constant. On the other hand, if we consider the time interval
between $t_{0}$ and the finite time $t_{0}+1/C$, we see that the scale factor
$b(t)$ goes to zero as $t\rightarrow t_{0}+1/C$ (see Fig. \ref{PlotEq10}b).
This could perhaps be interpreted as a sort of pre-inflationary period when,
immediately after the beginning of the universe, a dynamical compactification
of the extra dimensions takes place.

\begin{figure}[tbh]
\centering
\includegraphics[scale=0.6]{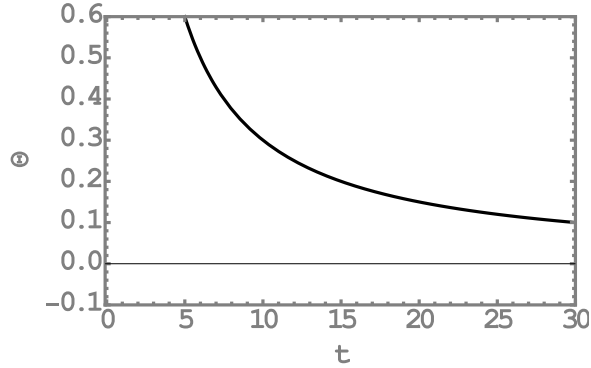} \caption{Expansion factor in
(\ref{ExpansFactor})}%
\label{PlotEq12}%
\end{figure}

If we now turn our attention to the expansion factor of the universe, a simple
calculation from (\ref{ds-model}) yields
\begin{equation}
\Theta_{(n)}=\frac{\dot{N}}{N}+6H_{a}+2(n-4)H_{b}.
\label{class expansion factor}%
\end{equation}
At this point, it should be mentioned that the expansion factor
(\ref{class expansion factor}) calculated for the solutions (\ref{class sol 1}%
) and (\ref{class sol 2}) is given by
\begin{equation}
\label{ExpansFactor}\Theta=\frac{3}{|C(t-t_{0})-1|}.
\end{equation}
That is, $\Theta_{(n)}$ has the same value and does not depend on the
dimension $n$. In both cases, we have expanding universes in which the
expansion rate decreases with time (see Figure \ref{PlotEq12}).

\FloatBarrier
\FloatBarrier

\begin{figure}[tbh]
\centering
\includegraphics[scale=0.7]{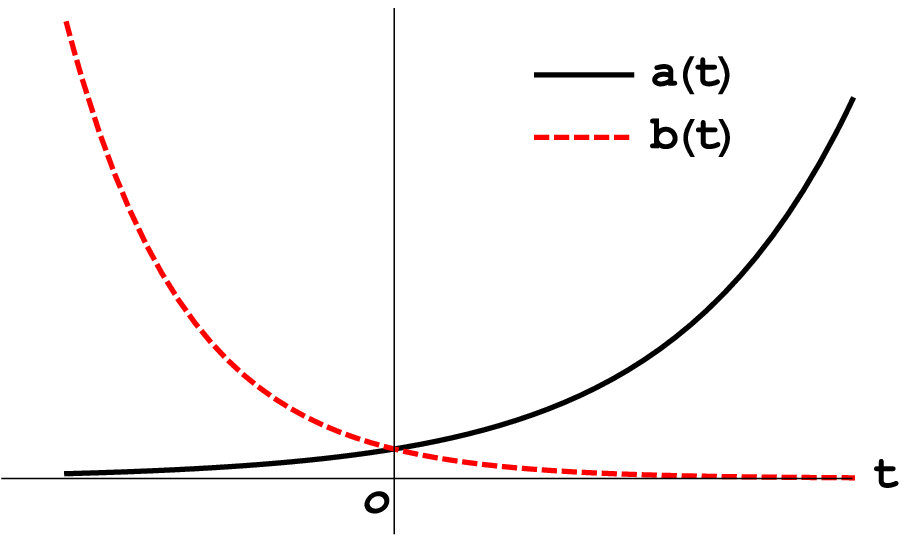} \qquad
\includegraphics[scale=0.7]{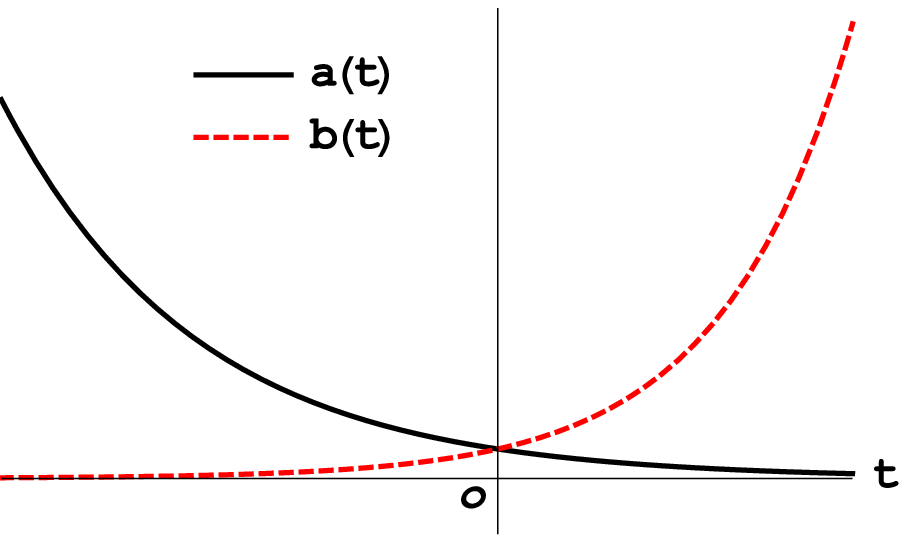} \caption{From left to right, scale
factors in (\ref{class sol}) and (\ref{class sol 0}) respectively}%
\label{PlotEq13}%
\end{figure}

Let us now consider two other different sets of solutions to the system of
equations (\ref{eq of motion}), which are given by
\begin{align}
\label{class sol}a(t) = a_{0} \exp\bigl[ \Lambda_{a} (t-t_{0})\bigr], \qquad
b(t) = b_{0} \exp\bigl[ - \Lambda_{b} (t-t_{0}) \bigr], \qquad\phi(t) =
\phi_{0}+D (t-t_{0}) ,
\end{align}
and
\begin{align}
\label{class sol 0}a(t) = a_{0} \exp\bigl[ -\Lambda_{a} (t-t_{0})\bigr],
\qquad b(t) = b_{0} \exp\bigl[ \Lambda_{b} (t-t_{0}) \bigr], \qquad\phi(t)
=\phi_{0}+ D (t-t_{0}),
\end{align}
where
\begin{align}
\label{solution constant}\Lambda_{a} = \sqrt{-\frac{ (n-4) D^{2} \omega
}{3(n-1)}} \qquad\text{and} \qquad\Lambda_{b} = \sqrt{-\frac{3 D^{2} \omega
}{(n-1)(n-4)} },
\end{align}
with $D$ being an integration constant, $\omega<0$, $a_{0}$, $b_{0}$, and
$\phi_{0}$ as defined above. Note that the solutions (\ref{class sol}) and
(\ref{class sol 0}) describe distinct scenarios. In the first, the
$4$-dimensional part of the universe is expanding while, the extra-
dimensional part is contracting. In the second, the dynamics of the universe
is reversed: the $4$-dimensional part collapses, while the extra dimensions
become larger (See Figure \ref{PlotEq13}). Moreover, as it is clear from
(\ref{class sol}) and (\ref{class sol 0}), in these universes there is no
space-time singularity. For the expansion factor, we have, from
(\ref{class expansion factor})
\[
\Theta_{(n)}=0,
\]
which means that, according to these models, the universe, as a whole, would
have no dynamics.

\FloatBarrier
\FloatBarrier

\subsection{The Hamiltonian formalism}

As we have already mentioned, the aim of this work is to investigate quantum
cosmological scenarios predicted by the geometrical scalar-tensor theory in
the case of anisotropic n-dimensional space-time. Following the methods of
canonical quantum cosmology, the first step is to carry out the canonical
quantization of the classical model. Thus let us compute the classical
Hamiltonian from the corresponding Lagrangian (\ref{Lagrang}).

It is not difficult to verify that the canonical momenta corresponding to the
variables $a,b$ and $\phi$ will be given, respectively, by
\begin{align}
P_{a}  &  =-\frac{12}{N}ab^{n-4}\dot{a}-\frac{6(n-4)}{N}a^{2}b^{n-5}\dot
{b},\nonumber\label{momentum}\\
P_{b}  &  =-\frac{2(n-4)(n-5)}{N}b^{n-6}a^{3}\dot{b}-\frac{6(n-4)}{N}%
b^{n-5}a^{2}\dot{a},\\
P_{\phi}  &  =\frac{2\omega}{N}a^{3}b^{n-4}\dot{\phi}.\nonumber
\end{align}
For $n>4$, the Hamiltonian takes the form
\begin{equation}
\mathcal{H}=\frac{N}{(n-2)ab^{n-6}}\left[  \frac{(n-5)}{12}\frac{{P_{a}}^{2}%
}{b^{2}}+\frac{1}{2(n-4)}\frac{{P_{b}}^{2}}{a^{2}}-\frac{P_{a}P_{b}}%
{2ab}+\frac{(n-2)}{4\omega a^{2}b^{2}}{P_{\phi}}^{2}\right]  .
\label{hamiltonian}%
\end{equation}
It turns out, however, that, the above form of $\mathcal{H}$ is not suitable
for working out the canonical quantization. \ A more convenient expression for
$\mathcal{H}$ can be obtained if we perform the following canonical
transformations: $A=\ln{a}$, $P_{A}=aP_{a}$, $B=\ln{b}$, $P_{B}=bP_{b}$,
$T=\frac{\phi}{P_{\phi}}$ and $P_{T}=\frac{{p_{\phi}}^{2}}{2}$
\cite{FarajollahiIntJTheorPhys49, FalcianoPhysRevD76}. The new Hamiltonian
written in terms of the new variables will be given by
\begin{equation}
\bar{\mathcal{H}}=\bar{N}\left[  \frac{\omega(n-5)}{6(n-2)}{P_{A}}^{2}
-\frac{\omega}{n-2}{P_{A}}{P_{B}}+\frac{\omega}{(n-2)(n-4)}{P_{B}}^{2}
+P_{T}\right]  , \label{transf hamiltonian}%
\end{equation}
with $\bar{N}=\frac{N}{2\omega a^{3}b^{n-4}}$, which, then, leads to the
equations of motion
\begin{align}
\dot{A}=\frac{\bar{N}\omega}{n-2}\left(  \frac{n-5}{3}P_{A}-P_{B}\right)
\qquad &  \text{,}\qquad\dot{P}_{A}=0,\nonumber\label{hamilton equations}\\
\dot{B}=\frac{\bar{N}\omega}{n-2}\left(  \frac{2}{n-4}P_{B}-\frac{P_{A}}%
{n-2}\right)  \qquad &  \text{,}\qquad\dot{P}_{B}=0,\\
\dot{T}=\bar{N}\qquad &  \text{and}\qquad\dot{P}_{T}=0.\nonumber
\end{align}
The solution of the above system (\ref{hamilton equations}) is easily
obtained, and is given by
\begin{align}
\label{hamiltonianeqssol}a(T)  &  =a_{0}\exp\left[  \frac{\omega}{n-2}\left(
\frac{n-5}{3}P_{A}-P_{B}\right)  T\right]  ,\nonumber\\
b(T)  &  =b_{0}\exp\left[  \frac{\omega}{n-2}\left(  \frac{2}{n-4}P_{B}%
-\frac{P_{A}}{n-2}\right)  T\right]  ,\\
\phi(T)  &  =\pm\sqrt{2P_{T}}T,\nonumber
\end{align}
with $P_{A}$, $P_{B}$ and $P_{T}$ being constants. As expected, one can easily
verify that (\ref{class sol 1}), (\ref{class sol 2}), (\ref{class sol}) and
(\ref{class sol 0}) are solutions of (\ref{hamilton equations}) when we set
$T\propto\phi$.

\section{The canonical quantization of the model}

\label{ModelQuant}

\subsection{The Wheeler-DeWitt equation}

In this section, we proceed with the quantization of the classical
cosmological model. By following the canonical quantization prescription
\begin{equation}
P_{A}\rightarrow-i\frac{\partial}{\partial A}\qquad\text{,}\qquad
P_{B}\rightarrow-i\frac{\partial}{\partial B}\qquad\text{,}\qquad
P_{T}\rightarrow-i\frac{\partial}{\partial T},
\end{equation}
the Wheeler-DeWitt equation
\begin{equation}
\hat{H}\Psi(A,B,T)=0, \label{wheeler DeWitt}%
\end{equation}
takes the form
\begin{equation}
\left\{  -\frac{\omega(n-5)}{6(n-2)}\frac{\partial^{2}}{\partial A^{2}}%
+\frac{\omega}{n-2}\frac{\partial^{2}}{\partial A\partial B}-\frac{\omega
}{(n-2)(n-4)}\frac{\partial^{2}}{\partial B^{2}}\right\}  \Psi(A,B,T)=i\frac
{\partial}{\partial T}\Psi(A,B,T), \label{schrodinger}%
\end{equation}
where $\hat{H}$ denotes the operator corresponding to $H$ (defined by
$\bar{\mathcal{H}}=\bar{N}H$ according to (\ref{transf hamiltonian})) and
$\Psi$ stands for the wave function of the universe. Clearly, Eq.
(\ref{wheeler DeWitt}) may be identified with the Schr\"{o}dinger equation
$\hat{H}\Psi=i\frac{\partial\Psi}{\partial T}$, where $T$ plays the role of
the parameter that measures the time evolution of the quantum system in
question. Let us just recall that the Hamiltonian $\hat{H}$ is required to be
a hermitian operator, with the usual inner product defined on $L^{2}$ as
\[
\left\langle \Psi_{1}\vphantom{\Psi_2}\right\vert \left.  \Psi_{2}%
\vphantom{\Psi_1}\right\rangle =\int_{-\infty}^{\infty}dA\int_{-\infty
}^{\infty}dB\Psi_{1}^{\ast}\Psi_{2},
\]
where are $\Psi_{1},\Psi_{2}$ are complex-valued measurable functions,
satisfying the boundary conditions
\[
\Psi(A\rightarrow\pm\infty)=0,\qquad\Psi(B\rightarrow\pm\infty)=0\qquad
(Dirichlet\;\;condition),
\]
or
\[
\frac{\partial\Psi}{\partial A}(A\rightarrow\pm\infty)=0,\qquad\frac{\partial\Psi}{\partial A}(B\rightarrow\pm\infty)=0,\qquad\frac
{\partial\Psi}{\partial B}(B\rightarrow\pm\infty)=0,\qquad\frac{\partial\Psi
}{\partial B}(A\rightarrow\pm\infty)=0\qquad(Neumann\;\;condition).
\]

At this point it is interesting to note that for $n=5$ the first term on the
left hand side in (\ref{transf hamiltonian}) vanishes and hence the
Schr\"{o}dinger equation takes the simple form
\begin{align}
\label{schrodinger five dim}\frac{\omega}{3} \left[  \frac{\partial^{2}%
}{\partial B^{2}} - \frac{\partial^{2}}{\partial A \partial B} \right]
\Psi(A,B,T) = - i \frac{\partial}{\partial T} \Psi(A,B,T).
\end{align}
Because of this great simplification we shall consider the case $n=5$
separately. The more general case corresponding to $n>5$ will be presented
next. For the sake of completeness, the quantization of the five-dimensional
model will be analyzed in Section \ref{appendix}.

\subsection{Solutions and expectation values for an n-dimensional quantum
universe}

\label{section2.3}

Since the Hamiltonian does not dependent on time explicitly we shall look for
stationary solutions of the form
\begin{align}
\label{wave function}\Psi(A,B,T)= \Phi(A,B) e^{-i ET},
\end{align}
where $E$ is a constant. As is well known, this leads to the time-independent
Schr\"{o}dinger equation
\begin{equation}
\hat{H}\Phi(A,B)=E\Phi(A,B).\nonumber
\end{equation}
If we now define new variables $u$ and $v$ by
\begin{align}
\label{transformation1}u= \sqrt{\frac{6(n-2)}{|\omega| (n-5)}} A + \sqrt
{\frac{(n-2)(n-4)}{|\omega|}} B \qquad v= \sqrt{\frac{6 (n-2)}{|\omega|(n-5)}}
A - \sqrt{\frac{(n-2)(n-4)}{|\omega|}} B,
\end{align}
then Eq.(\ref{schrodinger}) takes the form
\begin{align}
\label{transformed SEq}\left[  \eta^{(-)}\frac{\partial^{2}}{\partial u^{2}} +
\eta^{(+)} \frac{\partial^{2}}{\partial v^{2}} + E \right]  \Phi(u,v)=0,
\end{align}
where we have introduced the constants
\begin{align}
\eta^{(\pm)} = 2 \pm\sqrt{\frac{6(n-4)}{n-5}},\nonumber
\end{align}
and, for simplicity, we shall take $\omega>0$. To solve
Eq.(\ref{transformed SEq}) we write $\Phi(u,v)=\mathrm{U}(u)\mathrm{V}(v).$
This gives rise to the differential equations
\begin{align}
\frac{d^{2}\mathrm{U}}{du^{2}}+\frac{\lambda}{\eta^{(-)}}\mathrm{U}  &
=0,\nonumber\\
\frac{d^{2}\mathrm{V}}{dv^{2}}+\frac{E-\lambda}{\eta^{(+)}}\mathrm{V}  &  =0,
\end{align}
where $\lambda$ is a constant.

A particular solution to Eq.(\ref{transformed SEq}) will then easily be given
by
\begin{align}
\label{particular sol}\Phi_{\lambda, E}(\bar{u},\bar{v})= K \sin(\bar{u}
\sqrt{\lambda}) \sin(\bar{v} \sqrt{E+\lambda}),
\end{align}
where $\bar{u}=\frac{u}{\sqrt{|\eta^{(-)}|}}$ and $\bar{v}=\frac{v}{\sqrt
{\eta^{(+)}}}$ , $K$ is an arbitrary constant, and we are taking $\lambda>0$
and $E>-\lambda$. Clearly, the general solution to Eq.(\ref{schrodinger}) is
given by superposing the functions $\Psi_{\lambda,E}(\bar{u},\bar{v},T)$, that
is,
\begin{equation}
\Psi(\bar{u},\bar{v},T)=K\int_{0}^{\infty}dE_{1}\int_{0}^{\infty}dE_{2}%
F(E_{1},E_{2})e^{-i(E_{2}-E_{1})T}\sin(\bar{u}\sqrt{E_{1}})\sin(\bar{v}%
\sqrt{E_{2}}), \label{general solution}%
\end{equation}
where we are setting $E_{1}=\lambda$, $E_{2}=E+\lambda$, and $F(E_{1},E_{2})$
is a suitable weight function chosen to construct wave packets.

We are now going to choose a particular solution from (\ref{general solution})
by taking $F(E_{1},E_{2})=\exp\bigl[-\xi(E_{1}+E_{2})\bigr]$. It is not
difficult to verify that with this choice the normalized wave function reads
\begin{align}
\label{wave function pos omega}\Psi(\bar{u}, \bar{v}, T)=\sqrt{\frac
{\sqrt{3(n-2)(n-4)}}{\omega\pi}} \left( \frac{\xi}{\xi^{2} + T^{2}}\right)
^{3/2} \bar{u}\; \bar{v} \exp\left[ -\frac{1}{4} \left(  \frac{\bar{u}^{2}%
}{\xi-iT} + \frac{\bar{v}^{2}}{\xi+iT}\right) \right] .
\end{align}
In the same way, it is possible to obtain the wave function of the universe
from Eq.(\ref{schrodinger}) for $\omega<0$. In this case, a simple calculation
leads to
\begin{align}
\label{wave function neg omega}\Psi(\bar{u}, \bar{v}, T)=\sqrt{\frac
{\sqrt{3(n-2)(n-4)}}{|\omega| \pi}} \left( \frac{\xi}{\xi^{2} + T^{2}}\right)
^{3/2} \bar{u}\; \bar{v} \exp\left[ -\frac{1}{4} \left(  \frac{\bar{u}^{2}%
}{\xi+iT} + \frac{\bar{v}^{2}}{\xi-iT}\right) \right] .
\end{align}
Clearly, the wave function of the universe for $\omega>0$, given by Eq.
(\ref{wave function pos omega}), is just the complex conjugate of
(\ref{wave function neg omega}).

Let us now compute the expectation values $\left\langle a\right\rangle $ and
$\left\langle b\right\rangle $ of the scale factors $a(t)$ and $b(t)$
\footnote{We remark here that we shall adopt the many-worlds interpretation of
quantum mechanics \cite{Everett, Tipler}.}. Returning to the original
variables, we have, for any real $\omega$, that $\left\langle a\right\rangle $
will be given by
\begin{align}
\left<  a \right>  = \frac{|\omega|}{2} \sqrt{\frac{1}{3(n-2)(n-4)}}
\int^{\infty}_{-\infty} d\bar{u} \int^{\infty}_{-\infty} d\bar{v} \exp\left[
\frac{1}{2} \sqrt{\frac{|\omega|(n-5)}{6(n-2)}} \biggl(\sqrt{|\eta^{(-)}|}\;
\bar{u} + \sqrt{\eta^{(+)}}\; \bar{v} \biggr)\right]  |\Psi(\bar{u}, \bar
{v},T)|^{2},
\end{align}
which leads to
\begin{align}
\label{expectation value a}\left<  a \right>  = \frac{1}{8}\left[
\frac{\omega^{2}(n-5)}{36(n-2)} \Sigma^{2}(\xi,T^{2}) + \frac{4|\omega|}{n-2}
\sqrt{\frac{(n-4)(n-5)}{6}} \Sigma(\xi,T^{2}) + 8 \right]  \exp\left[
\frac{|\omega|}{4(n-2)} \sqrt{\frac{(n-4)(n-5)}{6}} \Sigma(\xi,T^{2})\right] ,
\end{align}
where here we have defined $\Sigma(\xi,T^{2})=\frac{\xi^{2}+T^{2}}{\xi}$. In a
similar manner, the expectation value $\left\langle b\right\rangle $\ of the
extra-dimensional scale factor $b(t)$ is given by
\begin{align}
\label{expectation value b}\left<  b \right>  = \frac{1}{8} \left[
\frac{\omega^{2}}{(n-2)(n-4)^{2}(n-5)} \Sigma^{2}(\xi,T^{2}) + \frac
{4|\omega|}{n-2} \sqrt{\frac{6}{(n-4)(n-5)}} \Sigma(\xi,T^{2}) + 8 \right]
\exp{ \left[ \frac{|\omega|}{4 (n-2)} \sqrt{\frac{6}{(n-4)(n-5)}} \Sigma
(\xi,T^{2})\right] }.
\end{align}

\FloatBarrier
\begin{figure}[tbh]
\centering
\includegraphics[scale=0.7]{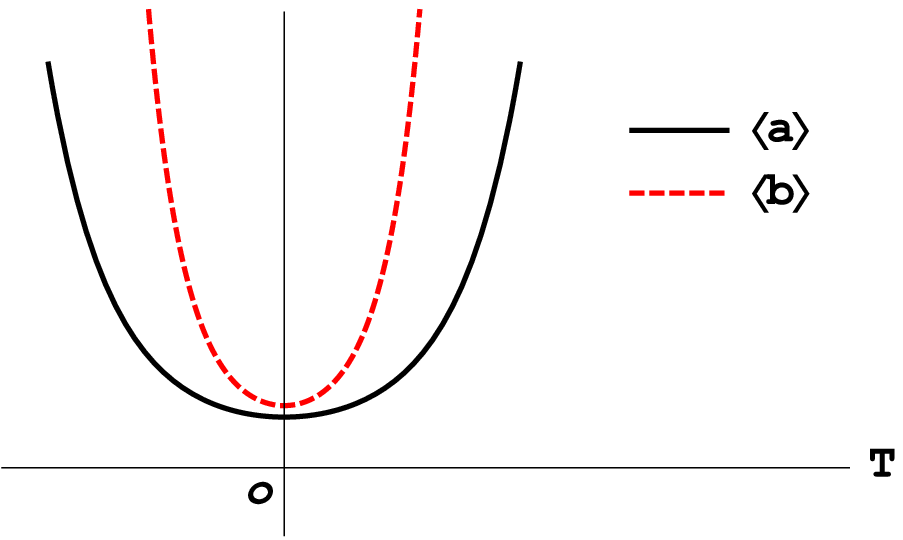} \qquad
\includegraphics[scale=0.7]{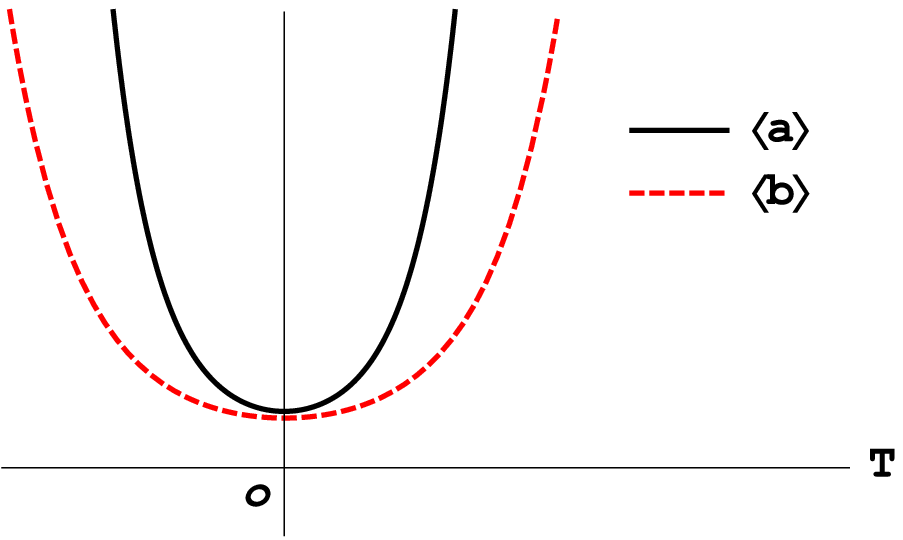}\caption{From left to right,
expectation values of the scale factors with n=6 and n=8 respectively}%
\label{Fig1}%
\end{figure}\FloatBarrier

An interesting point is, as was to be expected, that both expectation values
(\ref{expectation value a}) and (\ref{expectation value b}) coincide when
$n=7$. In Figure \ref{Fig1} the time behaviour of $\left\langle a\right\rangle
$ and $\left\langle b\right\rangle $\ is shown, qualitatively, for different
dimensions of the space-time. It should be mentioned that a similar picture
four a four-dimensional spacetime was obtained in
Ref.\cite{VakiliPhysLettB718}, where the exponentially decreasing (increasing)
classical solutions are replaced by scale factors of a bouncing universe.

From the expression of the expectation values given by the equations
(\ref{expectation value a}) and (\ref{expectation value b}), we get for the
expansion factor:
\begin{align}
\label{quantum exp factor}\Theta_{(n)}= \left[  6 \frac{1}{\left<  a \right> }
\frac{d \left<  a \right> }{dT} + 2 (n-4) \frac{1}{\left<  b \right> } \frac{d
\left<  b \right> }{dT} \right] \frac{dT}{dt} .
\end{align}

\begin{figure}[tbh]
\centering
\includegraphics[scale=0.6]{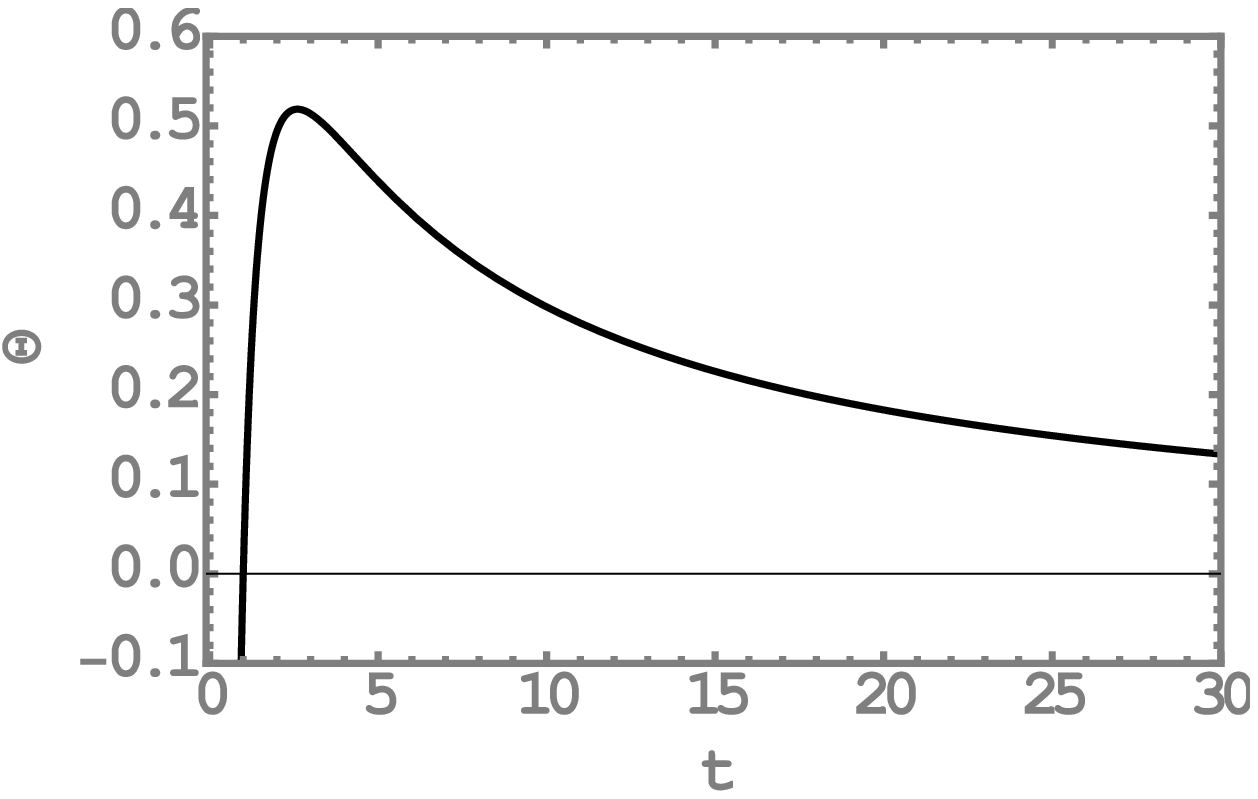}\qquad
\includegraphics[scale=0.6]{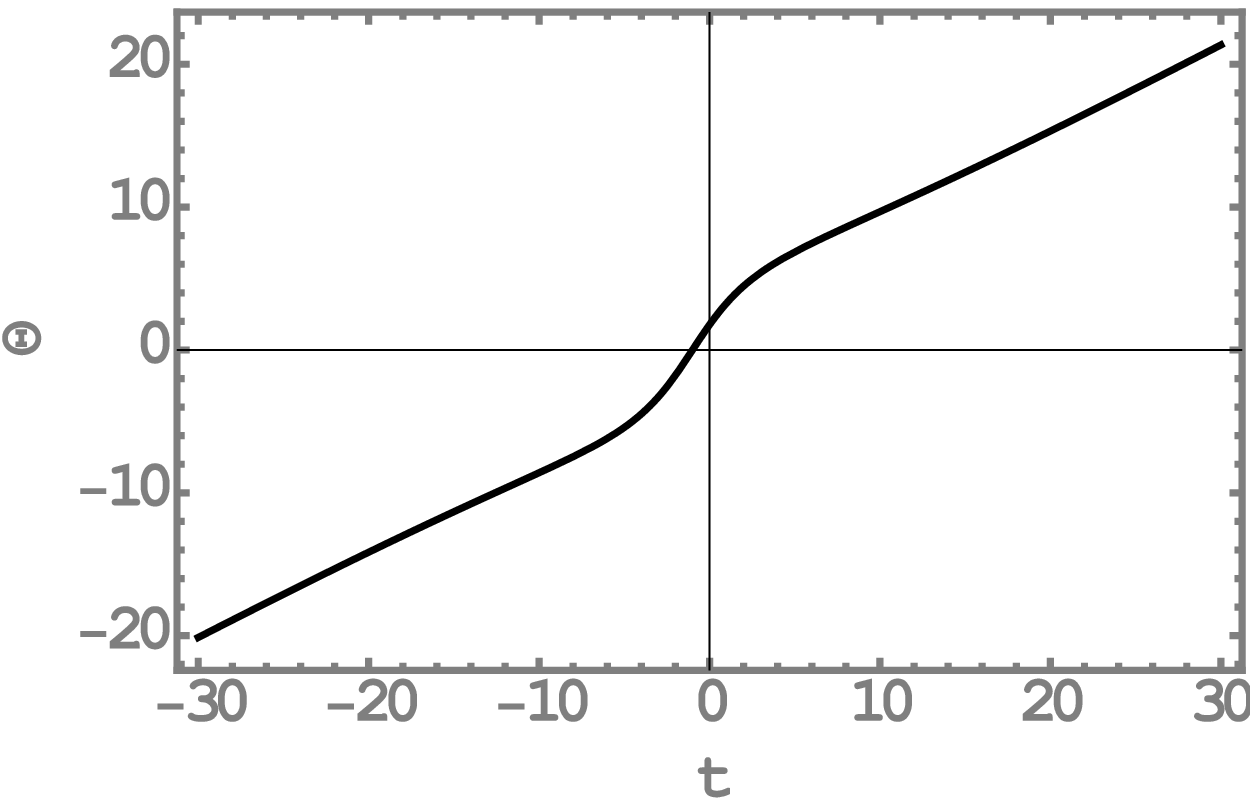}\caption{From left to right,
expansion factors corresponding to $T(t)$ related to the classical solutions
(\ref{class sol 1}) and (\ref{class sol 2}) respectively.}%
\label{Fig6}%
\end{figure}

Now let us consider the behaviour of the expansion factor for $n>5$, shown in
Figure \ref{Fig6}. It is important to highlight here that the behaviour of
$\Theta_{(n)}$, as given by (\ref{quantum exp factor}), does depend on the
space-time dimension, which is distinct from its behaviour at the classical
level. The expression of the evolution parameter $T$ as a function of time is
obtained from the solution of the Hamiltonian equations
(\ref{hamiltonianeqssol}) as $T=\frac{\phi(t)}{\sqrt{2 P_{T}}}$, being $P_{T}$
a constant and $\phi(t)$ is given by the classical model. As is expected in
the classical approximation, that is, when $t\rightarrow\infty$ and
$\xi\rightarrow0$, we recover the results obtained in Sec. (\ref{subsec2.1}).

\FloatBarrier
\FloatBarrier

\section{The quantization of the $5$-dimensional model}

\label{appendix}

In this section, we present the quantization of the model for $n=5$, taking
into consideration the solutions of Eq. (\ref{schrodinger five dim}).

By defining the variables
\[
x=2A+B\qquad y=B,
\]
and again applying the method of separation of variables,
Eq.(\ref{schrodinger five dim}) can be solved similarly to what was done in
Sec.(\ref{section2.3}). In this way, the normalized wave function of the
universe, for $\omega>0$, will be given by
\[
\Psi(x,y,T)=\frac{1}{2\sqrt{\pi}}\left(  \frac{\xi}{\xi^{2}+\frac{\omega^{2}%
}{9}T^{2}}\right)  ^{\frac{3}{2}}x\;y\exp\left\{  -\frac{1}{4}\left[
\frac{x^{2}}{\xi-i\frac{\omega}{3}T}+\frac{y^{2}}{\xi+i\frac{\omega}{3}%
T}\right]  \right\}  ,
\]
while, for $\omega<0$, we have
\begin{equation}
\Psi(x,y,T)=\frac{1}{2\sqrt{\pi}}\left(  \frac{\xi}{\xi^{2}+\frac{\omega^{2}%
}{9}T^{2}}\right)  ^{\frac{3}{2}}x\;y\exp\left\{  -\frac{1}{4}\left[
\frac{x^{2}}{\xi+i\frac{|\omega|}{3}T}+\frac{y^{2}}{\xi-i\frac{|\omega|}{3}%
T}\right]  \right\}  .
\end{equation}
The expectation value of the three-dimensional and extra-dimensional scale
factors will be given, respectively, by
\begin{align}
\label{expectation value a five dim}\left<  a \right>  = \frac{1}{4}\left[  1+
\frac{1}{4}\Sigma(\xi,T^{2})\right] ^{2} \exp\left[ \frac{\Sigma(\xi,T^{2}%
)}{4} \right] ,
\end{align}
\begin{align}
\label{expectation value b five dim}\left<  b \right>  = \frac{1}{4} \left[
1+ \Sigma(\xi,T^{2}) \right]  \exp\left[  \frac{\Sigma(\xi, T^{2})}{2} \right]
,
\end{align}
where we have defined $\Sigma(\xi, T^{2})=\frac{\xi^{2}+\frac{\omega^{2}}%
{9}T^{2}}{\xi}$. It follows, then, that the expansion factor for $n=5$
calculated from (\ref{expectation value a five dim}) and
(\ref{expectation value b five dim}) will be
\begin{align}
\label{expectation value five dim}\Theta_{(5)}= \frac{2 \omega^{2}}{9 \xi} T
\left\{  \frac{3 \left[ 12+ \Sigma(\xi,T^{2}) \right] }{2 \left[ 4+ \Sigma
(\xi,T^{2}) \right] } +\frac{3 + \Sigma(\xi,T^{2})}{1+ \Sigma(\xi,T^{2})}
\right\}  \frac{dT}{dt},
\end{align}
which exhibits the same profile shown in figure \ref{Fig6} and, as in the
previous cases, coincides with classical solutions as $\xi\rightarrow0$ and $t
\rightarrow\infty$.

\section{Final remarks}

\label{FinalRemarks}

In this work we have investigated the classical and quantum cosmological
scenarios predicted by a geometrical scalar-tensor gravitational theory, in an
anisotropic $n$-dimensional space-time. At the classical level, we have
obtained four different sets of solutions. Two of them represent a dynamical
singular universe bearing close resemblance to the well-known Kasner solution
\footnote{We note that the similarities with Kasner solution are due to the
form of the solutions for the scalar factors, since we are not considering the
vacuum case (we have effectively a scalar field stress tensor) and Kasner
condition is no longer present.}. The remaining sets of classical solutions
show an interesting picture. In one case we have a non-singular static
universe undergoing an expansion regime in the usual three dimensions, while
in the extra dimensions we have a contraction. We regard this result as some
kind of a $n$-dimensional generalization of the Chodos-Detweiler model
\cite{ChodosPhysRevD21}. The other case, leading to the opposite behaviour, in
which the role of the dimensions are reversed, is also allowed by the field equations.

At the quantum level, we have made use of the approach of quantum cosmology.
After carrying out a series of canonical transformations we obtained, after
applying the canonical quantization procedure, a Schr\"{o}dinger-like
differential equation for the wave function of the universe. We then found the
general solution to this equation and treated separately the cases $n=$ $5$
and $n>5$, which present similar behaviour. In the many-worlds interpretation
we found that the expectation values of the scale factors are clearly not
singular and, in fact, describe a bouncing universe. In other words, the
primordial cosmological singularity is avoided and the whole volume of the
universe undergoes a contraction phase, reaches a minimum volume and then
starts expanding. When compared with the classical regime, we could say that
at the quantum level the two classical solutions are linked to give rise to a
non-singular universe, in accordance with previous results
\citep{VakiliPhysLettB718}. 

To conclude, let us briefly comment on the role played by the Weyl field in
the framework of this geometrical scalar-tensor theory. As is already known,
the Weyl transformations preserve the geodesic lines and a number of other
geometrical objects, which then implies the physical equivalence of the class
of Weyl frames, at the classical level \cite{Almeida}. It is possible to show
that there exists a class of Weyl transformations which induce canonical
transformations in the reduced Hamiltonian of the original action
\cite{Barreto}. However, we still do not know how to extend this classical equivalence to the quantum level, if this is possible at all \cite{Anderson}.

Finally, we would like to remark that, with regard to the well-known problem
of time in quantum cosmology, it seems appealing to consider that the
geometrical nature of the scalar field may lead to a more natural
identification of this field with the time parameter that governs the
evolution of the quantum variables.


\section*{Acknowledgments}

We thank F. Dahia and N. Pinto-Neto for fruitful discussions and suggestions.
Thanks also go to CAPES and CNPq for financial support. We thank the referees
for useful and relevant comments that helped to improve the quality
of the manuscript.



\end{document}